\pdfoutput=1
\documentclass[aps,superscriptaddress,twocolumn,showpacs,showkeys,amsmath,amsthm,amsfonts,amssymb,nofootinbib]{revtex4}
\usepackage{graphicx}% Include figure files

\allowdisplaybreaks[0]

\begin{document}

%\preprint{APS/123-QED}
%\newcommand{\binom}[2]{\genfrac{(}{)}{0pt}{}{#1}{#2}}
%$\binom{A}{B}$

\title{Generalized Maxwell-Boltzmann, Bose-Einstein, Fermi-Dirac and Acharya-Swamy Statistics and the P\'olya Urn Model}% Force line breaks with \\
%Generalized Maxwell-Boltzmann, Bose-Einstein and Fermi-Dirac Probability Distributions and Cross-Entropy Measures

\author{Robert K. Niven}
\email{r.niven@adfa.edu.au}
\affiliation{School of Aerospace, Civil and Mechanical Engineering, The University of New South Wales at ADFA, Northcott Drive, Canberra, ACT, 2600, Australia.
}
\affiliation{Niels Bohr Institute, University of Copenhagen, Copenhagen \O, Denmark.}

\author{Marian Grendar}
\email{marian.grendar@savba.sk}
\affiliation{Department of Mathematics, Faculty of Natural Sciences, Bel University, Tajovskeho 40, 974 01 Banska Bystrica, Slovakia.
%; Institute of Mathematics and Computer Science of the Slovak Academy of Sciences (SAS); Institute of Measurement Science of SAS.
}

\date{15 August 2008}% can use \today

\begin{abstract}
Generalized probability distributions for Maxwell-Boltzmann, Bose-Einstein and Fermi-Dirac statistics, with unequal source probabilities $q_i$ for each level $i$, are obtained by combinatorial reasoning. For equiprobable degenerate sublevels, these reduce to those given by Brillouin in 1930, more commonly given as a statistical weight for each statistic. These distributions and corresponding cross-entropy (divergence) functions are shown to be special cases of the P\'olya urn model, involving neither independent nor identically distributed (``ninid'') sampling. The most probable P\'olya distribution contains the Acharya-Swamy intermediate statistic.

\end{abstract}


%Use showkeys class option if keyword

\maketitle

%% ############################################################################
\section{\label{intro}Introduction}
%% ############################################################################
Over 80 years ago, many physical systems were found to be governed by Bose-Einstein (BE) or Fermi-Dirac (FD) statistics \cite{Bose_1924, Einstein_1924, Einstein_1925, Fermi_1926, Dirac_1926}, as distinct from classical degenerate Maxwell-Boltzmann (MB) statistics developed in the 19th century \cite{Boltzmann_1877, Planck_1901}. The combinatorial weights of these statistics - the number of configurations (microstates) of each identifiable realization (complexion, macrostate) of the system - are widely given as \citep[e.g.][]{Bose_1924, Einstein_1924, Einstein_1925, Fermi_1926, Dirac_1926, Brillouin_1927, Brillouin_1930, Tolman_1938, Brillouin_1951b, Davidson_1962}:
\begin{align}
\mathbb{W}_{MB}  &= N!\prod\limits_{i = 1}^s {\frac{{g_i ^{n_i } }}{{n_i !}}},
\label{eq:W_MB} \\
%\end{equation}
%\begin{equation}
\mathbb{W}_{BE}  &= \prod\limits_{i = 1}^s {\frac{{(g_i  + n_i  - 1)!}}{{(g_i  - 1)!n_i !}}}, 
\label{eq:W_BE} \\
%\end{equation}
%\begin{equation}
\mathbb{W}_{FD}  &= \prod\limits_{i = 1}^s {\frac{{g_i !}}{{n_i !(g_i  - n_i )!}}},  
\label{eq:W_FD}
\end{align}
where $i$ denotes each distinguishable level (outcome or state) of the system, from a total of $s$ levels; $n_i$ is the number of entities in each level $i$; $N= \sum\nolimits_{i = 1}^s {n_i} $ is the total number of entities; and $g_i$ is the number of distinguishable sublevels within each level $i$ (the degeneracy or multiplicity). In each case, the dimensionless entropy per entity $H$ is obtained from the Boltzmann principle \cite{Boltzmann_1877, Planck_1901}:
\begin{equation}
H = \frac{S_{total}}{kN} = \frac{{\ln \mathbb{W}}}{N},
\label{eq:Boltzmann1}
\end{equation}
where $S_{total}$ is the total dimensional (thermodynamic) entropy and $k$ is the Boltzmann constant.  Maximizing the entropy defined by \eqref{eq:Boltzmann1} (``MaxEnt''), subject to the constraints on the system, is therefore equivalent to identifying the system with its most probable realization, a method of inductive reasoning which can be termed the {\it maximum probability} (``MaxProb'') principle \cite{Boltzmann_1877, Planck_1901, Vincze_1972, Grendar_G_2001, Niven_CIT, Niven_2005, Niven_2006, Niven_MaxEnt07, Niven_EntropyJ08}. Normally, \eqref{eq:Boltzmann1} is subject to the asymptotic limit $N \to \infty$ (simplistically using the Stirling approximation, $\ln m! \approx m\ln m - m$, or more rigorously with Sanov's \cite{Sanov_1957} theorem), to give the entropy function for each statistic \cite{Boltzmann_1877, Planck_1901, Bose_1924, Einstein_1924, Einstein_1925, Fermi_1926, Dirac_1926, Brillouin_1927, Brillouin_1930, Tolman_1938, Brillouin_1951b, Davidson_1962}. Recently, \eqref{eq:Boltzmann1} has been applied directly to give the {\it exact} or {\it non-asymptotic} entropy function for each statistic \cite{Niven_2005, Niven_2006, Niven_MaxEnt07, Niven_EntropyJ08}, enabling the application of statistical mechanics to systems containing small numbers of entities. The non-asymptotic BE and FD entropy functions have peculiar information-theoretic properties, with important implications for quantum mechanics \cite{Niven_2005, Niven_2006}. 

The above weights were generalized by Brillouin [\onlinecite{Brillouin_1930}, also in \onlinecite{Fortet_1977, Read_1983}] to give:
\begin{align}
\mathbb{P}_{MB|G} &= \frac{N!}{G^N} \prod\limits_{i = 1}^s {\frac{{g_i ^{n_i } }}{{n_i !}}} ,
\label{eq:Pu_MB} \\
%\end{equation}
%\begin{equation}
\mathbb{P}_{BE|G} &= \frac{{N!(G - 1)!}}{{(G + N - 1)!}}\prod\limits_{i = 1}^s {\frac{{(g_i  + n_i  - 1)!}}{{n_i !(g_i  - 1)!}}},
\label{eq:Pu_BE} \\
%\end{equation}
%\begin{equation}
\mathbb{P}_{FD|G} &= \frac{{N!(G - N)!}}{{G!}}{\rm{ }}\prod\limits_{i = 1}^s {\frac{{g_i !}}{{n_i !(g_i  - n_i )!}} }.
\label{eq:Pu_FD} 
\end{align}
where $\mathbb{P}$ denotes a correctly normalized probability distribution and $G= \sum\nolimits_{i = 1}^s {g_i}$ is the total degeneracy.  These are given in many references \citep[e.g.][]{Feller_1957} as the probabilities $s^{-N} N! /\prod\nolimits_{i = 1}^s {n_i!}$, ${\binom{s+N-1}{N}} ^{-1}$ and ${\binom{s}{N}} ^{-1}$ respectively, i.e.\ without considering degenerate sublevels within each level.  In contrast, the following generalized distributions were given by Kapur \cite{Kapur_1989a}:
\begin{align}
\mathbb{P}_{BE}^{Kapur} &= \prod\limits_{i = 1}^s {\frac{{(g_i  + n_i  - 1)!}}{{n_i !(g_i  - 1)!}}\frac{{q_i ^{n_i } }}{{(1 + q_i )^{g_i  + n_i } }}} ,
\label{eq:P_BE_Kapur} \\
%\end{equation}
%\begin{equation}
\mathbb{P}_{FD}^{Kapur} &= \prod\limits_{i = 1}^s {\frac{{g_i !}}{{n_i !(g_i  - n_i )!}}q_i ^{n_i } (1 - q_i )^{g_i  - n_i } }.
\label{eq:P_FD_Kapur}
\end{align}
where $q_i$ is the source probability\footnote{In the MaxEnt community, $q_i$ is termed the {\it prior probability}, implying a connection to Bayesian inference; the existence and form of such a connection is the subject of much debate \citep[e.g.][]{Knuth_2005, Caticha_MaxEnt07, Grendar_2008}.} of an entity in level $i$. In the limits $q_i \! \ll \! 1$ and $g_i \! \gg \! n_i$, these imply:
\begin{equation}
\mathbb{P}_{MB}^{Kapur} = \prod\limits_{i = 1}^s {\frac{{g_i ^{n_i } q_i ^{n_i } }}{{n_i !}}},
\label{eq:P_MB_Kapur}
\end{equation}
 (see later discussion). The Brillouin and Kapur sets of distributions are incompatible, and their range of validity and interrelationships have not been explored.

Of great interest in recent years is the possible occurrence of ``anyon'' particles, intermediate between bosons and fermions.  Several intermediate statistics have been proposed.  A long-standing philosophical approach is to consider the allocation of indistinguishable particles to distinguishable degenerate boxes, with a maximum of $d$ particles per box; the end-members $d \to \infty$ and $d=1$ respectively give BE and FD statistics \cite{Gentile_1940, Guenault_M_1962, Kapur_K_1992}.  A different statistic by Haldane and Wu \cite{Haldane_1991, Wu_1994} emerges from a generalized Pauli exclusion principle, involving coupled interactions between particles in occupancies $\{n_i\}$. A third approach was given as an ansatz by Acharya and Swamy \cite{Acharya_S_1994}, and later justified (approximately) by quantum symmetry \cite{Acharya_S_2004}, or by exclusion \cite{Wu_1994, Polychronakos_1996, Ilinskaya_etal_1996}, particle kinetics \cite{Kaniadakis_etal_1996} or fractional entanglement \cite{Zhou_2000} arguments.

The aims of this work are:\ (i) to report the correct, generalized probability distribution for MB, BE and FD statistics containing source terms $q_i$, and thereby to examine the validity of \eqref{eq:W_MB}-\eqref{eq:W_FD} and \eqref{eq:Pu_MB}-\eqref{eq:P_MB_Kapur}; (ii) to examine these as special cases of the P\'olya statistic, involving ``neither independent nor identically distributed'' (``ninid'') sampling \cite{Eggenberger_P_1923, Polya_1931, Steyn_1951, Patil_etal_1984, Patil_R_1985, Johnson_K_1997}; and (iii) to report the asymptotic and non-asymptotic cross-entropy (divergence) functions and minimum cross-entropy (MinXEnt) distributions for P\'olya systems based on the MaxProb principle \cite{Grendar_N_2007}.  It is shown that the generalized P\'olya statistic contains, and therefore justifies, the Acharya-Swamy \cite{Acharya_S_1994} intermediate statistic. 

%%############################################################################
\section{\label{deriv}Derivations}
%% ############################################################################
%
\begin{figure}[t]
%\begin{center}
\setlength{\unitlength}{0.6pt}
  \begin{picture}(410,565)
   \put(0,0){\small (c)}
   \put(5,15){\includegraphics[width=80mm]{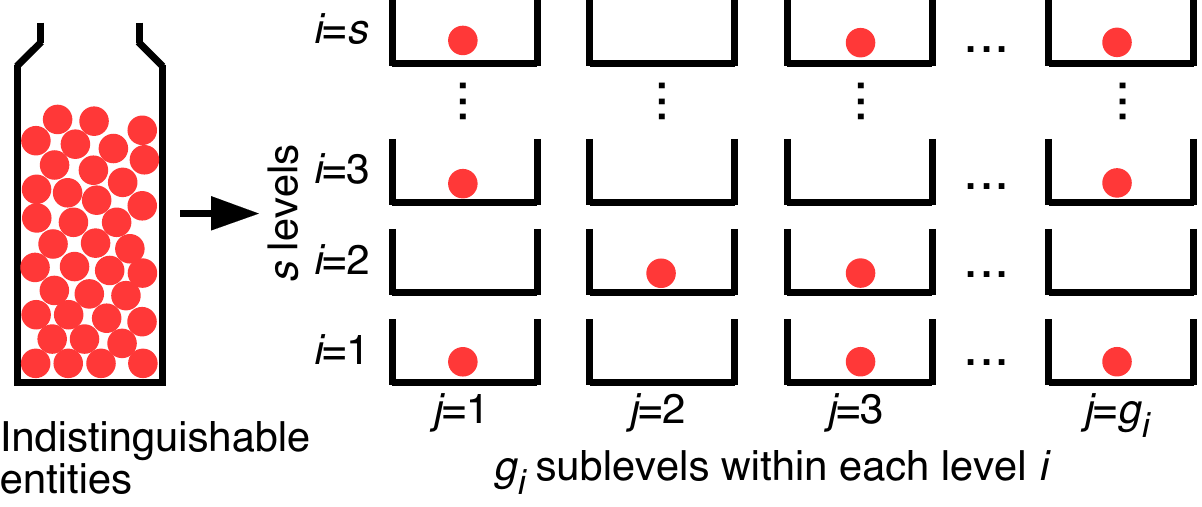} }
   \put(0,190){\small (b)}
   \put(5,205){\includegraphics[width=80mm]{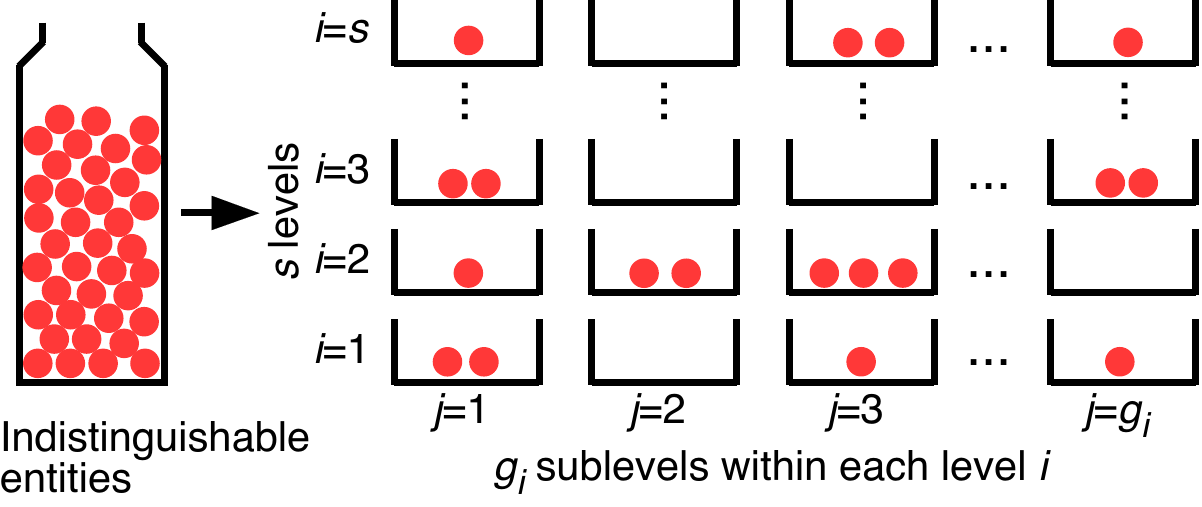} }
   \put(0,380){\small (a)}
   \put(5,395){\includegraphics[width=80mm]{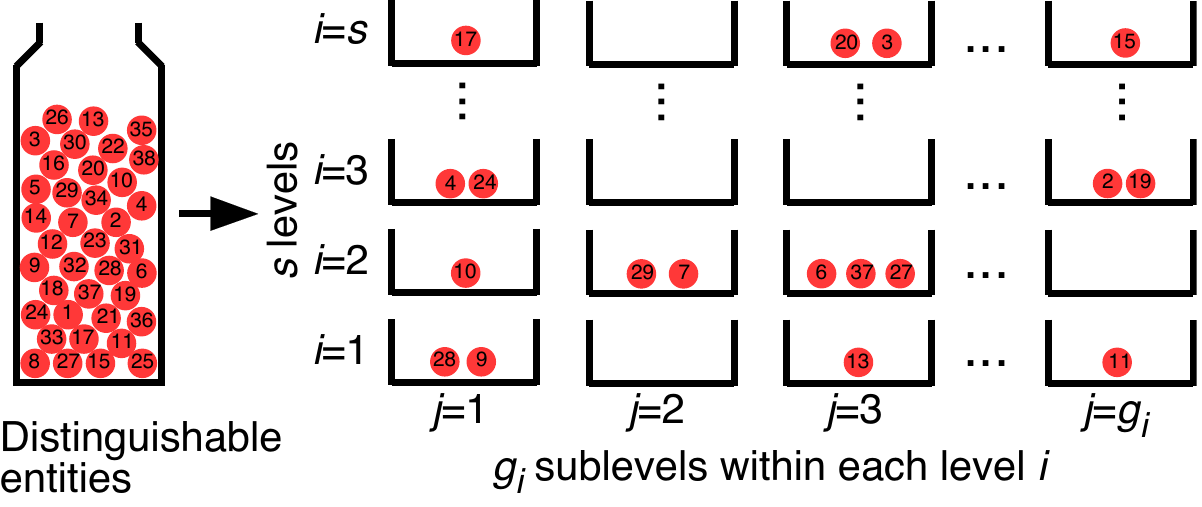} }
  \end{picture}
%\end{center}
\caption{Configurations of (a) Maxwell-Boltzmann, (b) Bose-Einstein and (c) Fermi-Dirac ball-in-box models, in which all $N$ entities are fully allocated (the $g_i$ need not be equal). }
\label{fig:comb1}
\end{figure}
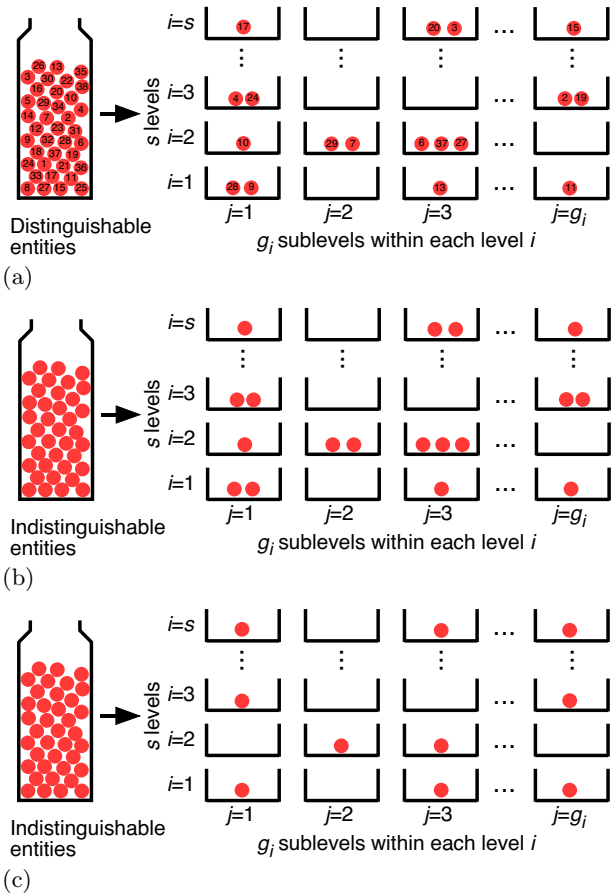

By enumeration of several simple examples:\ (a) a non-degenerate system; (b) a uniformly degenerate system (equal $g_i$); (c) a non-uniformly degenerate system; and (d) systems with unequal $g_i$ and $q_i$, it is readily verified that the Brillouin distributions (\ref{eq:Pu_MB})-(\ref{eq:Pu_FD}) - and implicitly, the above weights (\ref{eq:W_MB})-(\ref{eq:W_FD}) - are valid only for the case of equal source probabilities for each degenerate sublevel, $q_i=g_i/G$.  They do not apply to equal source probabilities for each level, $q_i=s^{-1}$, or to more general forms. The Kapur distributions (\ref{eq:P_BE_Kapur})-(\ref{eq:P_MB_Kapur}) do not satisfy normalization, $\sum\nolimits_{\mathbb{P} \in \{\mathbb{P}\}} \mathbb{P}=1$. It is therefore necessary to reexamine the probabilistic basis of each statistic. 

{\it Maxwell-Boltzmann statistics}: Firstly, consider a system in which $N$ distinguishable balls (entities) are arranged amongst $s$ distinguishable boxes (levels), in which each level contains $g_i$ distinguishable sublevels, as shown in Figure \ref{fig:comb1}a.  Each level has the source probability $q_i$ of being filled (not necessarily equal), with $\sum\nolimits_{i = 1}^s {q_i}=1$. We consider each {\it realization} of the system, containing $n_i$ balls in each level $i$, $\forall i$; this in turn is a set of {\it configurations}, or distinguishable arrangements of balls in boxes. In counting configurations, we distinguish between rearrangements of balls between boxes, but do not (or need not) consider rearrangements of balls within each box (see Figure \ref{fig:comb1}a). No account is made of the order of allocating balls to boxes.  Of course, the realizations can be chosen in many different ways; the occupancies $\{n_i\}$ are selected here for their utility in physical systems.

Clearly, the probability of an individual arrangement of $n_i$ balls in the $i$th level is $q_i^{n_i}$, and so the probability of any specified configuration {\it amongst the levels} is $\prod\nolimits_{i = 1}^s {q_i ^{n_i }}$.  To obtain $\mathbb{P}_{MB}$, this must be multiplied by the number of ways in which the balls can be rearranged amongst the levels to give the same realization, $N! /\prod\nolimits_{i = 1}^s {n_i!}$. The result is the {\it non-degenerate} multinomial distribution:
\begin{equation}
\mathbb{P}_{MB} = N! \prod\limits_{i = 1}^s { \frac{q_i ^{n_i }}{n_i !} }.
\label{eq:P_MB}
\end{equation}
Another viewpoint is that if the source probability of the $i$th level is $q_i$, that of the $j$th sublevel of the $i$th level is $q_{ij}=q_i/g_i, \forall j=1,...,g_i$, assuming equiprobable filling of sublevels within a level. The probability of a configuration {\it amongst the sublevels} - i.e.\ of $n_{ij}$ balls in the $j$th sublevel of the $i$th level, $\forall \{i,j\}$ - is then $\prod\nolimits_{i = 1}^s {\prod\nolimits_{j = 1}^{g_i} {(q_i/g_i)}^{n_{ij} }}$ = $\prod\nolimits_{i = 1}^s  {(q_i/g_i)}^  {\sum\nolimits_{j = 1}^{g_i}{n_{ij} }  }$ = $\prod\nolimits_{i = 1}^s {(q_i/g_i) ^{n_i }}$.  This must be multiplied by the number of configurations amongst sublevels, $N! \prod\nolimits_{i = 1}^s {{g_i}^{n_i}/{n_i!}}$, again giving \eqref{eq:P_MB}. 
%The validity of \eqref{eq:P_MB} is also verified by enumeration of the examples considered (Appendix G).  
For equiprobable sublevels $q_{ij} \! =\! G^{-1}$, $q_i=g_i/G$ and \eqref{eq:P_MB} reduces to the Brillouin form \eqref{eq:Pu_MB}.  Since the total number of configurations amongst sublevels is $G^N$, this is proportional to the degenerate weight \eqref{eq:W_MB}.

\begin{figure}[t]
%\begin{center}
\setlength{\unitlength}{0.6pt}
  \begin{picture}(410,170)
   \put(5,0){\includegraphics[width=80mm]{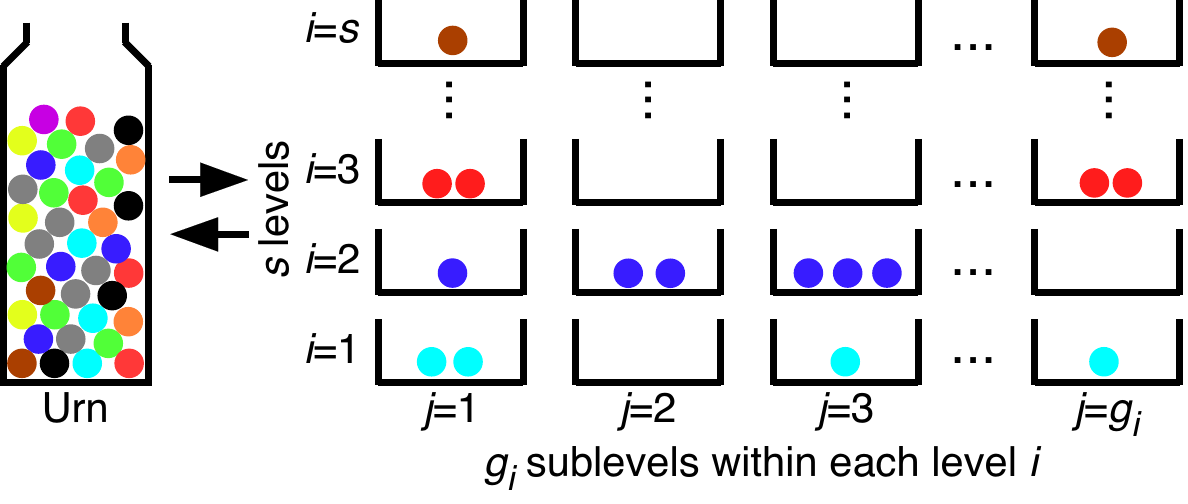} }
  \end{picture}
%\end{center}
\caption{An urn model, in which balls are drawn, recorded and returned to an urn, until $N$ balls have been sampled.}
\label{fig:urn}
\end{figure}

An important conclusion is that when considering a generalized source probability $q_i$ unrelated to the degeneracy $g_i$, the latter becomes irrelevant to the probability of each configuration. We can in fact put $q_i=m_i/M$, where $m_i \in \mathbb{N}$ is a fictitious degeneracy controlled by the source probability and $M= \sum\nolimits_{i = 1}^s {m_i}$, whence:
\begin{equation}
\mathbb{P}_{MB} = N! \prod\limits_{i = 1}^s {\frac{{(m_i/M) ^{n_i } }}{{n_i !}}} ,
\label{eq:P_MB_M} 
\end{equation}
Eq.\ (\ref{eq:P_MB_M}) has a strong connection to {\it urn models} of probability distributions \citep[e.g.][]{Feller_1957, Berg_1988, Johnson_K_1997}, as shown in Figure \ref{fig:urn}, in which balls are drawn from an urn containing $M$ balls, made up of $m_i$ balls of  each level (color) $i$.  In each sampling event, a ball is drawn from the urn, its color recorded, and is returned to the urn in accordance with some rule, until a sample of $N$ balls has been recorded. In the MB case, the balls are drawn {\it with replacement}, hence the $q_i$ remain constant.  As shown in Figure \ref{fig:urn}, the balls are indistinguishable except for their color; the weight $N! /\prod\nolimits_{i = 1}^s {n_i!}$ now arises from the number of ways in which a given set of balls can be sampled, i.e.\ the number of ordered {\it sequences} of balls which make up each realization or {\it type}.  Urn models thus capture many features of sets of alphabetic characters $a_i$ drawn from an alphabet $\mathcal{A}=\{a_i\}$, considered in information theory \cite{Shannon_1948, Cover_T_2006}. The urn model insight is of benefit to later analyses.

The above distribution can be generalized to the case of unequal sources $q_{ij}$ for each sublevel, to give:
\begin{equation}
\mathbb{P}_{MB}^{'} = N! \prod\limits_{i = 1}^s {\frac{ g_i ^{n_i }  } {n_i !}}  \prod\limits_{j = 1}^{g_i} {q_{ij}^{n_{ij}} } ,
\label{eq:P_MB_double} 
\end{equation}
with normalization $ \sum\nolimits_{i=1}^{s} \sum\nolimits_{j=1}^{g_i}  q_{ij}=1$. It is unclear why such a complicated arrangement might be needed, since each sublevel could be handled as a separate level; but it does give a recipe for ``coarse graining'' if this is desired.

{\it Bose-Einstein statistics}: We now consider the same balls-in-boxes problem but with indistinguishable balls, as shown in Figure \ref{fig:comb1}b.  The standard argument is that there are ${\binom{g_i+n_i-1}{n_i}}$ distinguishable ways of arranging $n_i$ indistinguishable balls in $g_i$ boxes, $\forall i$, hence the number of configurations amongst sublevels in a realization is given by the BE weight \eqref{eq:W_BE}. For equally probable configurations, the governing probability is obtained by dividing by the total number of configurations amongst sublevels, ${\binom{G+N-1}{N}}$, giving the Brillouin form \eqref{eq:Pu_BE}. 

We now employ the urn model analogy, to consider the generalized BE distribution:
\begin{equation}
\mathbb{P}_{BE} = \frac{{N!(M - 1)!}}{{(M + N - 1)!}}\prod\limits_{i = 1}^s {\frac{{(m_i  + n_i  - 1)!}}{{n_i !(m_i  - 1)!}}}.
\label{eq:P_BE_M} 
\end{equation}
Using $q_i=m_i/M$ and parameter $\beta^{-}=N/M \le 1$ gives:
\begin{equation}
\mathbb{P}_{BE} = \frac{{N!(N/\beta^{-} - 1)!}}{{(N/\beta^{-} + N - 1)!}}\prod\limits_{i = 1}^s {\frac{{(q_i N/\beta^{-}  + n_i  - 1)!}}{{n_i !(q_i N/\beta^{-}  - 1)!}}}, 
\label{eq:P_BE_beta} 
\end{equation}
with $q_i N/ \beta^{-} \! \in \! \mathbb{N}$. This is a variant of the {\it multivariate negative hypergeometric} or {\it Dirichlet compound multinomial distribution} \cite{Johnson_K_1969, Patil_etal_1984, Jensen_1985}, as is the Brillouin BE distribution \eqref{eq:Pu_BE} itself.  This arises from a complicated scenario of {\it sampling without replacement}, in which two sets of balls are drawn from an urn; the conditional distribution of the second sample, given the first, is the one of interest \cite{Patil_etal_1984, Jensen_1985}. For equiprobable sublevels, $q_i=g_i/G$; taking $m_i=g_i$ and $M=G$, \eqref{eq:P_BE_M} reduces to the Brillouin form \eqref{eq:Pu_BE}, and thus is proportional to the BE weight \eqref{eq:W_BE}. For equiprobable levels $q_i=s^{-1}$, thus taking $m_i=1$ and $M=s$, it yields the non-degenerate BE distribution ${\binom{s+N-1}{N}} ^{-1}$. Outside these two cases, \eqref{eq:P_BE_beta} gives a generalized BE distribution with unequal source probabilities $q_i$, which satisfies normalization and several other requirements of a governing probability distribution (see later sections). 

{\it Fermi-Dirac statistics}: Here we also have indistinguishable balls, but each sublevel contains a maximum of one ball, as shown in Figure \ref{fig:comb1}c.  By standard reasoning, there are ${\binom{g_i}{n_i}}$ distinguishable ways of arranging $n_i \in \{0,1\}$ indistinguishable balls in $g_i$ boxes, $\forall i$; thus the number of configurations amongst sublevels in a realization is given by the FD weight \eqref{eq:W_FD}. For equiprobable configurations, the total number of configurations amongst sublevels is ${\binom{G}{N}}$, giving the Brillouin distribution \eqref{eq:Pu_FD}. 

Applying the urn model analogy gives the generalized FD distribution:
\begin{equation}
\mathbb{P}_{FD} = \frac{{N!(M - N)!}}{{M!}} \prod\limits_{i = 1}^s {\frac{{m_i !}}{{n_i !(m_i  - n_i )!}} },
\label{eq:P_FD_M} 
\end{equation}
with the condition $m_i \ge n_i$. In terms of $q_i$ and parameter $\beta^{+}=N/M \le 1$:
\begin{equation}
\mathbb{P}_{FD} = \frac{{N!(N/\beta^{+} - N)!}}{{(N/\beta^{+})!}} \prod\limits_{i = 1}^s {\frac{{(q_i N/\beta^{+})!}}{{n_i !(q_i N/\beta^{+}  - n_i )!}} },
\label{eq:P_FD_beta}
\end{equation}
with $q_i N/ \beta^{+} \! \in \! \mathbb{N}$ and $q_i N/ \beta^{+} \!  \ge n_i$. This and the Brillouin FD distribution \eqref{eq:Pu_FD} are variants of the {\it multivariate hypergeometric distribution} \cite{Johnson_K_1969, Patil_etal_1984, Jensen_1985}, which arises from a simple case of {\it sampling without replacement} \cite{Johnson_K_1969, Patil_etal_1984, Jensen_1985}.  For equiprobable sublevels, this function also reduces to the Brillouin form \eqref{eq:Pu_FD}, usually expressed as the FD weight \eqref{eq:W_FD}, whilst for equiprobable levels, it gives the non-degenerate FD distribution ${\binom{s}{N}}^{-1}$. Aside from these cases, \eqref{eq:P_FD_beta} gives a generalized FD distribution with unequal source probabilities $q_i$. 

%%############################################################################
\section{\label{Disc}Discussion}
For $a, b \in \mathbb{N}$, the following two inequalities are well known:
\begin{gather}
{\frac{{a !}}{{(a  - b )!}}} \le a^b \le {\frac{{(a  + b  - 1)!}}{{(a  - 1)! }}},  
\label{eq:lemma}
\end{gather}
with equality for $a \gg b$. Proofs are given in Appendix B. These and the weights \eqref{eq:W_MB}-\eqref{eq:W_FD} yield the inequalities \citep[e.g.][]{Davidson_1962}:
\begin{align}
\mathbb{W}_{FD}  \le \frac {\mathbb{W}_{MB}}{N!} \le \mathbb{W}_{BE}
\label{eq:W_compare}
\end{align}
with equality as $g_i \gg n_i, \forall i$.  If we consider the Brillouin distributions \eqref{eq:Pu_MB}-\eqref{eq:Pu_FD}, one obtains:
\begin{gather}
\begin{split}
\mathbb{P}_{FD|G}  \gtreqqless  {\mathbb{P}_{MB|G}} \gtreqqless \mathbb{P}_{BE|G}
\end{split}
\label{eq:Pu_compare}
\end{gather}
The variable direction arises from the terms which converge to $\prod \nolimits_{i=1}^s{ g_i^{n_i}}/ G^N$, for which opposite examples can be found (see Appendix B); equality in \eqref{eq:Pu_compare} occurs when $G \gg N$ and $g_i \gg n_i, \forall i$.  

Much attention has been paid to \eqref{eq:W_compare}, especially to the term ${\mathbb{W}_{MB}}/{N!}$, which has even been labelled a separate statistic for ``corrected Boltzon'' particles \cite{Fowler_1936, Davidson_1962}. However, comparing \eqref{eq:W_compare}-\eqref{eq:Pu_compare}, it is evident that the $N!$ divisor in \eqref{eq:W_compare} arises from the fact that the MB weight is normalized differently to the BE and FD weights. Since probabilities are always normalized consistently, no such discrepancy appears in the Brillouin distributions \eqref{eq:Pu_MB}-\eqref{eq:Pu_FD}, nor indeed in the generalized forms \eqref{eq:P_MB}, \eqref{eq:P_BE_beta} and \eqref{eq:P_FD_beta} derived herein. In consequence, the ``corrected Boltzon'' statistic has no independent physical meaning, being merely an artefact of the use of weights; indeed, except in the asymptotic limit $N \to \infty$, it gives absurd results.  This demonstrates an important point, that probabilistic inference using the MaxProb principle should {\it always} be based directly on true probabilities $\mathbb{P}$; any contrary method (e.g., one based on weights $\mathbb{W}$) can lead to logical inconsistencies \cite{Niven_MaxEnt07}.

%############################################################################
\section{\label{Polya}The P\'olya Distribution and MaxProb}
We now consider the P\'olya urn model (Figure \ref{fig:urn}), in which a ball is sampled from an urn, recorded, and returned to the urn; in addition, $c \in \mathbb{Z}$ balls of the same color (level) are also added to the urn. The probability of drawing the realization $\{n_i\}$ from the urn after $N$ draws is \cite{Eggenberger_P_1923, Polya_1931, Steyn_1951, Patil_etal_1984, Patil_R_1985, Johnson_K_1997}:
\begin{gather}
\mathbb{P}_{Polya} = \frac{N!}{\prod_{i=1}^s n_i
!} \frac{\prod_{i=1}^s m_i(m_i+c)\dotsm(m_i+(n_i-1)c)} {M(M+c)\dotsm(M+(N-1)c)},
\label{eq:Polya_distrib}
\end{gather}
The P\'olya urn thus describes a simple, analytic case of ``neither independent nor identically distributed'' (``ninid'') sampling, of which the bivariate form is widely used in studies of biological contagion \citep [e.g.][]{Patil_R_1985}.  

The P\'olya model \eqref{eq:Polya_distrib} can be written in the following forms, valid respectively for $c=0$, $c>0$ and $c<0$:
\begin{align}
\mathbb{P}_{\substack{ Polya\\ c=0}} &= \frac {N!}{M^N} \prod\limits_{i = 1}^s {\frac{{m_i^{n_i } }}{{n_i !}}} ,
\label{eq:Polya_distrib0}
\\
\mathbb{P}_{\substack{ Polya\\ c>0}} &= \frac{N!}{\prod_{i=1}^s n_i
!} \frac{\Gamma\left(\frac{M}{c}\right)}{\Gamma\left(\frac{M}{c} +
N\right)} \prod_{i=1}^s \frac{\Gamma\left(\frac{m_i}{c} +
n_i \right)}{\Gamma\left(\frac{m_i}{c}\right)},
\label{eq:Polya_distrib1}
\\
\mathbb{P}_{\substack{ Polya\\ c<0}} &=  \frac{N !}{\prod_{i=1}^s n_i !} \frac{\Gamma\left(1-\frac{M}{c}- N\right)} {\Gamma\left(1 -
\frac{M}{c}\right)}\prod_{i=1}^s \frac{\Gamma\left(1
-\frac{m_i}{c}\right)}{\Gamma\left(1- \frac{m_i}{c} - n_i\right)}.
\label{eq:Polya_distrib2}
\end{align}
These distributions encompass the MB, BE and FD governing distributions \eqref{eq:P_MB_M}, \eqref{eq:P_BE_M} and \eqref{eq:P_FD_M}.  Simplified forms of \eqref{eq:Polya_distrib} and \eqref{eq:Polya_distrib1}-\eqref{eq:Polya_distrib2} were given by \cite{Ilinskaya_etal_1996, Kaniadakis_etal_1996}.  In light of the P\'olya model, we can set aside the complicated urn model for the multivariate negative hypergeometric distribution (the generalized BE statistic \eqref{eq:P_BE_M}), to offer the following explanations for each statistic:
\begin{list}{$\bullet$}{\topsep 2pt \itemsep 2pt \parsep 0pt \leftmargin 8pt \rightmargin 0pt \listparindent 0pt
\itemindent 0pt}
\item MB statistic: P\'olya urn with c=0 (sampling with replacement); 
\item BE statistic: P\'olya urn with c=1 (sampling with double replacement); 
\item FD statistic: P\'olya urn with c=-1 (sampling without replacement); 
\end{list}
It is curious that the last two representations are equivalent to the allocation of indistinguishable balls to distinguishable boxes (with a further restriction in the FD case).  Whilst the full implications of this insight are not clear, it does provide an alternative philosophical foundation for the construction of quantum statistics. 

The P\'olya distributions \eqref{eq:Polya_distrib1}-\eqref{eq:Polya_distrib2} can be further analyzed using an extension of Boltzmann's principle \eqref{eq:Boltzmann1}:
\begin{equation}
D = - \frac{{\ln \mathbb{P}}}{N},
\label{eq:Boltzmann3}
\end{equation}
where $D$ is the generalized cross-entropy (directed divergence or negative relative entropy) function defined by the MaxProb principle.  Applying \eqref{eq:Boltzmann3} in the asymptotic limit $N \to \infty$ to \eqref{eq:Polya_distrib0}-\eqref{eq:Polya_distrib2}, using $p_i=n_i/N$, $q_i=m_i/M$ and $\beta=N/M$, gives:
\begin{align}
-D_{\substack{ Polya\\ c=0}} &=  \sum\limits_{i = 1}^s -p_i \ln \frac {p_i}{q_i} 
\label{eq:D_Polya0_st} 
\displaybreak[0] 
\\
\begin{split}
-D_{\substack{ Polya\\ c>0}} &=  \sum\limits_{i = 1}^s \biggl\lbrace - p_i \ln p_i + p_i \Bigl( \frac{1}{\beta c} \Bigr) \ln \Bigl( \frac{1}{\beta c} \Bigr)
\\& \mspace{-60.0mu} 
- p_i \Bigl(\frac{1}{\beta c}+1\Bigl) \ln \Bigl(\frac{1}{\beta c}+1 \Bigl)   
\\& \mspace{-60.0mu} 
+ \Bigl(\frac{q_i}{\beta c}+p_i \Bigl) \ln \Bigl(\frac{q_i}{\beta c}+p_i \Bigl) - \Bigl(\frac{q_i}{\beta c} \Bigl) \ln \Bigl(\frac{q_i}{\beta c} \Bigl) \biggr\rbrace ,
\end{split}
\label{eq:D_Polya1_st} 
\displaybreak[0] 
\\
\begin{split}
-D_{\substack{ Polya\\ c<0}} &=  \sum\limits_{i = 1}^s \biggl\lbrace - p_i \ln p_i 
- p_i \Bigl( - \frac{1}{\beta c} \Bigr) \ln \Bigl( - \frac{1}{\beta c} \Bigr) 
\\& \mspace{-60.0mu} 
+ p_i \Bigl(-\frac{1}{\beta c}-1\Bigl) \ln \Bigl(-\frac{1}{\beta c}-1 \Bigl)   
\\ & \mspace{-60.0mu} 
- \Bigl(-\frac{q_i}{\beta c}-p_i \Bigl) \ln \Bigl(-\frac{q_i}{\beta c}-p_i \Bigl) + \Bigl(- \frac{q_i}{\beta c} \Bigl) \ln \Bigl(-\frac{q_i}{\beta c} \Bigl) \biggr\rbrace.
\end{split}
\label{eq:D_Polya2_st}
\end{align}
with restriction $|c|<q_i/\beta p_i, \forall i$ (whence $|c|<m_i/n_i, \forall i$) for $c<0$.  Note \eqref{eq:D_Polya0_st} is the Kullback-Liebler cross-entropy \cite{Kullback_L_1951}.  Proof of these results based on Sanov's theorem is given elsewhere \cite{Grendar_N_2007}; a more simplistic calculation may be conducted by applying Stirling's approximation to all factorials and gamma functions. Both forms of the P\'olya cross-entropy \eqref{eq:D_Polya1_st}-\eqref{eq:D_Polya2_st} have been shown to satisfy many important information-theoretic properties, including non-negativity, lower semicontinuity, convexity, partition inequality and the conditional limit theorem \cite{Grendar_N_2007}.

Applying the MaxProb principle, the relevant cross-entropy  \eqref{eq:D_Polya0_st}, \eqref{eq:D_Polya1_st} or \eqref{eq:D_Polya2_st} can be minimized (MinXEnt) subject to the following constraints:
\begin{gather}
\sum\limits_{i = 1}^s {p_i }= 1.
\label{eq:C0}
\\
\sum\limits_{i = 1}^s {p_i f_{ri} }= \left\langle {f_r } \right\rangle, \quad r = 1,...,R,
\label{eq:Cr}
\end{gather}
where $f_{ri}$ is the value of the function $f_r$ in the $i$th level and $\left\langle {f_r} \right\rangle$ is its mathematical expectation. Expremization by the Lagrangian method gives the inferred ``most probable" distribution for each statistic:
\begin{align}
p_{\substack{ Polya,i\\ c=0}}^{*} &=  q_i e^{  - \lambda_{0}^{c=0} - \sum\nolimits_{r=1}^R {\lambda_r  f_{ri}} },
\label{eq:p_Polya0_st} 
\\
p_{\substack{ Polya,i\\ c>0}}^{*} &= \frac {q_i/\beta c} { e^{ \lambda_{0}^{c>0} + \sum\nolimits_{r=1}^R {\lambda_r  f_{ri}} } - 1},
\label{eq:p_Polya1_st} 
\\
p_{\substack{ Polya,i\\ c<0}}^{*} &= - \frac {q_i/\beta c} { e^{  \lambda_{0}^{c<0} + \sum\nolimits_{r=1}^R {\lambda_r  f_{ri}} } + 1},
\label{eq:p_Polya2_st} 
\end{align}
where $\lambda_{r}, r=0,...,R$ are Lagrangian multipliers, with constants absorbed into the $\lambda_{0}$ terms. For equiprobable sublevels $q_i=g_i/G$, sample size $\beta=N/G$ and the above values of $c$, the cross-entropy functions \eqref{eq:D_Polya0_st}-\eqref{eq:D_Polya2_st} and distributions \eqref{eq:p_Polya0_st}-\eqref{eq:p_Polya2_st} reduce to those given by Brillouin \cite{Brillouin_1927, Brillouin_1930}; these in turn simplify - up to constant factors - to the commonly used entropy functions and distributions obtained from the weights\footnote{Physicists typically consider constraints on normalization \eqref{eq:C0} and energy $f_{1i}=E_i$, $\langle f_1 \rangle=\langle E \rangle$ in \eqref{eq:Cr}, giving multipliers $\lambda_1=1/kT$ and $\lambda_0 = - \mu /kT$, where $k$ is the Boltzmann constant, $T$ is absolute temperature and $\mu$ can be viewed as a chemical potential.} \eqref{eq:W_MB}-\eqref{eq:W_FD} \cite{Bose_1924, Einstein_1924, Einstein_1925, Fermi_1926, Dirac_1926, Brillouin_1927, Brillouin_1930, Tolman_1938, Brillouin_1951b, Davidson_1962}.

For $N \ll \infty$, the non-asymptotic cross-entropy functions and distributions for the above systems can also be derived using Boltzmann's principle \eqref{eq:Boltzmann3}, as reported in Appendix A. These encompass the non-asymptotic MB, BE and FD entropy functions and distributions reported previously \cite{Niven_2005, Niven_2006}. 

To summarize, generalized MB, BE and FD statistics, containing unequal source probabilities for each level, can be obtained using the combinatorial definition of cross-entropy \eqref{eq:Boltzmann3} in conjunction with probability distributions generated using urn models. All such cases emerge as special instances of the P\'olya urn model, involving a simple scenario of urn modification during sampling.  Of the systems examined, only the MB system in the asymptotic limit yields the Kullback-Leibler cross-entropy \eqref{eq:D_Polya0_st} \cite{Kullback_L_1951}; in consequence, the latter is {\it not} universal in application, and {\it cannot} be used to infer the most probable realization of BE, FD or P\'olya systems except in special limiting cases (nor even of the MB system away from the asymptotic limit $N \to \infty$). This conclusion contradicts the dominant viewpoint in information theory, in which the Shannon entropy \cite{Shannon_1948} and Kullback-Liebler cross-entropy \cite{Kullback_L_1951} are considered to be the paramount functions for inductive reasoning (``statistical inference''), based on their axiomatic and information-theoretic definitions \citep[c.f.][]{Jaynes_1957, Jaynes_2003}.  This conclusion is, however, a necessary consequence of Boltzmann's principle.

Eqs.\ \eqref{eq:p_Polya0_st}-\eqref{eq:p_Polya2_st} can be amalgamated as:
\begin{align}
p_{Polya,i}^{*} &= \frac {q_i} { e^{ \lambda_{0}' + \sum\nolimits_{r=1}^R {\lambda_r  f_{ri}} } - \beta c},
\label{eq:p_Polya_int_st} 
\end{align}
%where $\lambda_{0}'$ incorporates further constant terms. 
This can be compared to the ansatz presented by Acharya and Swamy \cite{Acharya_S_1994}, in the present notation:
\begin{align}
p_{AS,i}^{*} &=  \frac {g_i/N} { e^{ \lambda_{0}' + \sum\nolimits_{r=1}^R {\lambda_r  f_{ri}} } - a},
%\frac{n_{AS,i}^{*}}{N} =
\label{eq:p_AS_int_st} 
\end{align}
in which $a=0,1$ and $-1$ respectively give MB, BE and FD statistics, and $-1 \le a \le 1$ their intermediate interpolation. The P\'olya representation \eqref{eq:p_Polya_int_st} thus contains the Acharya-Swamy intermediate statistic \eqref{eq:p_AS_int_st} as a special case.  It therefore justifies this statistic using a ``ninid'' urn sampling scheme, an explanation rather different to that of previous studies \cite{Wu_1994, Polychronakos_1996, Ilinskaya_etal_1996, Kaniadakis_etal_1996, Zhou_2000, Acharya_S_2004}.

%%############################################################################
\section{Conclusions}

Generalized probability distributions for MB, BE and FD statistics, with unequal source probabilities $q_i$ for each level $i$, are derived by extension of the distributions given by Brillouin \cite{Brillouin_1927, Brillouin_1930}, using an urn model analogy. These are shown to be special instances of the P\'olya urn model, involving a simple scenario of urn modification during sampling (``ninid'' sampling).  The resulting cross-entropy functions and MinXEnt distributions are derived by Boltzmann's principle (the MaxProb method). The general form of the P\'olya distribution contains the Acharya-Swamy \cite{Acharya_S_1994} intermediate statistic, as an exact result without approximation.

We also show that the ``corrected Boltzon'' statistic based on $\mathbb{W}_{MB}/N!$ has no independent meaning, but is an artefact of the use of weights instead of probabilities in Boltzmann's principle. Furthermore, since the P\'olya cross-entropy function in general differs from the Kullback-Liebler cross-entropy \cite{Kullback_L_1951}, the latter does not give the most probable realization of BE, FD or P\'olya systems except in special limiting cases.

As a final comment, the statistic of Haldane and Wu \cite{Haldane_1991, Wu_1994} considers intermediacy as a consequence of changes in occupancies $\{n_i\}$ due to interactions.  The Acharya-Swamy-P\'olya model considers intermediacy due to changes in the source probabilities $\{q_i\}$ - hence in the degeneracies $\{g_i\}$ - during sampling. Either approach is justifiable on physical grounds; their common features and possible synthesis deserve further investigation. 

%% ############################################################################
%%%%%%%%%%%%%%%%%%%%%

\begin{acknowledgments}
The first author was supported by a Marie Curie Incoming International Fellowship project 039729, awarded by the European Commission under Framework Programme 6. The second author was supported by VEGA grant 1/3016/06 and Australian Research Council grant no. DP0210999; the hospitality of the School of Computer Science and Engineering, The University of New South Wales, Australia (with special thanks to Arthur Ramer) and of the Niels Bohr Institute, University of Copenhagen, Denmark, are gratefully acknowledged.

\end{acknowledgments}

%%############################################################################
\section{Appendix A: Non-Asymptotic Distributions}

The non-asymptotic forms of the P\'olya cross-entropy functions can be obtained directly from Boltzmann's principle \eqref{eq:Boltzmann3}, without taking the asymptotic limit $N \to \infty$. Such statistics are required for systems of small numbers of non-interacting entities \cite{Niven_2005, Niven_2006, Niven_MaxEnt07, Niven_EntropyJ08}.  From \eqref{eq:Boltzmann3} and the governing distributions \eqref{eq:P_MB}, \eqref{eq:P_BE_beta} and \eqref{eq:P_FD_beta}, the functions are:
\begin{align}
-D_{\substack{ Polya\\ c=0}}^{(N)} &= \frac{1}{N} \sum\limits_{i = 1}^s {\left\lbrace p_i \ln N! + p_i N \ln q_i - \ln (p_i N) ! \right\rbrace } 
\label{eq:D_Polya0_N} 
\displaybreak[0] \\
\begin{split}
-D_{\substack{ Polya\\ c>0}}^{(N)} &= \frac{1}{N} \sum\limits_{i = 1}^s \biggl\lbrace  p_i \ln \frac{ N! \; \Gamma(N/\beta c )} {\Gamma(N/\beta c + N)}  \\
& \quad + \ln \frac{ \Gamma (q_i N/\beta c  + p_i N ) } {(p_i N)! \; \Gamma (q_i N/\beta c )}  \biggr\rbrace
\end{split}
\label{eq:D_Polya1_N} 
\displaybreak[0] \\
\begin{split}
-D_{\substack{ Polya\\ c<0}}^{(N)} &= \frac{1}{N} \sum\limits_{i = 1}^s \biggl\lbrace  p_i \ln \frac{N! \; \Gamma(1 - N/\beta c - N)}{ \Gamma(1-N/\beta c)}  \\
& \quad  + \ln \frac{\Gamma(1-q_i N/\beta c)} {(p_i N) ! \; \Gamma(1-q_i N/\beta c  - p_i N ) } \biggr\rbrace
\end{split}
\label{eq:D_Polya2_N} 
\end{align}
in which constant terms are brought inside the sum using \eqref{eq:C0}.  Applying MinXEnt subject to \eqref{eq:C0}-\eqref{eq:Cr} gives the inferred ``most probable'' distribution for each statistic:
\begin{align}
p_{\substack{ Polya,i\\ c=0}}^{\#} &=  \frac{1}{N} \Lambda^{-1} \left\lbrace  \frac{1}{N} \ln N! + \ln q_i - \lambda_0^{c=0} - \sum\limits_{r=1}^R {\lambda_r  f_{ri}} \right\rbrace
\label{eq:p_Polya0_N} 
\displaybreak[0] 
\\
\begin{split}
p_{\substack{ Polya,i\\ c>0}}^{\#} &= \frac{1}{N} \Lambda^{-1} \biggl\lbrace  \frac{1}{N}  \ln \frac{ N! \; \Gamma(N/\beta c )} {\Gamma(N/\beta c + N)}  
\\
&  + \Lambda \bigl( q_i N/\beta c  + p_{\substack{ Polya,i\\ c>0}}^{\#} N  - 1 \bigl) - \lambda_0^{c>0} - \sum\limits_{r=1}^R {\lambda_r  f_{ri}} \biggr\rbrace
\end{split}
\label{eq:p_Polya1_N} 
\displaybreak[0] \\
\begin{split}
p_{\substack{ Polya,i\\ c<0}}^{\#} &= \frac{1}{N} \Lambda^{-1} \biggl\lbrace  \frac{1}{N}  \ln \frac{N! \; \Gamma(1-N/\beta c - N)}{\Gamma(1-N/\beta c)}   
\\
&   - \Lambda \bigl( -q_i N/\beta c  - p_{\substack{ Polya,i\\ c<0}}^{\#} N \bigl) - \lambda_0^{c<0} - \sum\limits_{r=1}^R {\lambda_r  f_{ri}} \biggr\rbrace
\end{split}
\label{eq:p_Polya2_N} 
\end{align}
where $\lambda_r, r=0,...,R$ again are Lagrangian multipliers, $\Lambda(x)=\psi(x+1)=y$ is a shifted digamma function, defined for convenience, and $\Lambda^{-1}(y)=\psi^{-1}(y)-1$ is its inverse. None of these functions allow explicit factorization of a partition function $Z=e^{\lambda_0}$; the last two forms are also implicit in $p_i^{\#}$.

%%############################################################################
\section{Appendix B: Inequalities}
For $b >0$ the terms in \eqref{eq:lemma} can be written:
\begin{align}
\frac{a !}{(a  - b )!} 
%&=  \frac{ a (a-1) (a-2) ... 1} { (a-b) (a-b-1) ... 1}
&= a (a-1) ... (a-b+1)
\label{eq:ineq1a}
\\
\frac{(a+b-1) !}{(a  - 1 )!} 
%&=  \frac{ (a+b-1)(a+b-2) ... 1} { (a-1) (a-2) ... 1}
&= (a+b-1)(a+b-2) ... a
\label{eq:ineq2a}
\end{align}
As $a/b \to \infty$, each factor in \eqref{eq:ineq1a}-\eqref{eq:ineq2a} will approach $a$ respectively from below or above. Both forms thus converge to $a^b$ in accordance with \eqref{eq:lemma}; \eqref{eq:W_compare} follows.  $\square$

From \eqref{eq:ineq1a}-\eqref{eq:ineq2a}, both $\mathbb{P}_{BE|G}$ and $\mathbb{P}_{FD|G}$ converge to $\mathbb{P}_{MB|G}$. However, the sign of the inequalities can vary. This is illustrated by the set of realizations for the Brillouin case $N=4$, $s=3$ and $g_i=3, \forall i$, for which the probabilities \eqref{eq:Pu_MB}-\eqref{eq:Pu_FD} are listed in Table \ref{tab:mult} (some realizations are excluded from the FD set).  Depending on the realizations, the order can be $\mathbb{P}_{FD|G} < \mathbb{P}_{MB|G} < \mathbb{P}_{BE|G}$, $\mathbb{P}_{FD|G} > \mathbb{P}_{MB|G} > \mathbb{P}_{BE|G}$ or $\mathbb{P}_{FD|G} < \mathbb{P}_{MB|G} > \mathbb{P}_{BE|G}$.  In this and many other examples calculated, we did not observe $\mathbb{P}_{FD|G} > \mathbb{P}_{MB|G} < \mathbb{P}_{BE|G}$, although this result is not proven.

\begin{table}[h]
\begin{tabular}{p{0pt} p{51pt} p{38pt} p{14pt} p{38pt} p{14pt} p{35pt}  } 
\hline
&Realization $[n_1,n_2,n_3]$ &$\mathbb{P}_{FD}$	&$\gtreqqless$ &$\mathbb{P}_{MB}$	&$\gtreqqless$  &$\mathbb{P}_{BE}$	 \\
\hline
&[1, 1, 2]	&0.21429	&$>$	&0.14815	&$>$	&0.10909\\
&[1, 2, 1]	&0.21429	&$>$	&0.14815	&$>$	&0.10909\\
&[2, 1, 1]	&0.21429	&$>$	&0.14815	&$>$	&0.10909\\
&[2, 2, 0]	&0.07143	&$<$	&0.07407	&$>$	&0.07273\\
&[2, 0, 2]	&0.07143	&$<$	&0.07407	&$>$	&0.07273\\
&[0, 2, 2]	&0.07143	&$<$	&0.07407	&$>$	&0.07273\\
&[1, 3, 0]	&0.02381	&$<$	&0.04938	&$<$	&0.06061\\
&[1, 0, 3]	&0.02381	&$<$	&0.04938	&$<$	&0.06061\\
&[3, 1, 0]	&0.02381	&$<$	&0.04938	&$<$	&0.06061\\
&[3, 0, 1]	&0.02381	&$<$	&0.04938	&$<$	&0.06061\\
&[0, 1, 3]	&0.02381	&$<$	&0.04938	&$<$	&0.06061\\
&[0, 3, 1]	&0.02381	&$<$	&0.04938	&$<$	&0.06061\\
&[4, 0, 0]	&NA	&	&0.01235	&$<$	&0.03030\\
&[0, 4, 0]	&NA	&	&0.01235	&$<$	&0.03030\\
&[0, 0, 4]	&NA	&	&0.01235	&$<$	&0.03030\\
\hline
\end{tabular}
\caption []{Probabilities of each realizations for $N=4$, $s=3$ and $g=[3,3,3]$ (NA=not applicable).}
\label{tab:mult}
\end{table}

%% ############################################################################

\end{document}